\documentclass[10pt]{iopart}
\usepackage{amssymb}
\usepackage{url}
\usepackage{graphicx}
\usepackage{subfigure}
\usepackage{citesort}
\begin{document}
\title[Large-deviation properties of power grids]
{Large-deviation properties of resilience of power grids}

\author{Timo Dewenter and Alexander K Hartmann}
\address{Institut f\"ur Physik, Carl von Ossietzky Universit\"at 
Oldenburg, 26111 Oldenburg, Germany}
\ead{timo.dewenter@uni-oldenburg.de}

\begin{abstract}
We study the distributions of the resilience of power flow models 
against transmission line failures
via a so-called backup capacity. We consider three 
ensembles of random networks and in addition, the topology of the British 
transmission
power grid. The three ensembles are Erd\H{o}s-R\'enyi 
random graphs, Erd\H{o}s-R\'enyi random graphs with 
a fixed number of links, and spatial networks where the nodes are embedded 
in a two dimensional plane. We investigate numerically the 
probability density functions (pdfs) down to the tails to gain 
insight in very resilient and 
very vulnerable networks. This is achieved via large-deviation techniques 
which allow us to study very rare values which occur with 
probability densities below $10^{-160}$. We find that the 
right tail of the 
pdfs towards larger backup capacities follows an exponential 
with a strong curvature. This is confirmed by the rate function which 
approaches a limiting curve for increasing network sizes. Very resilient 
networks are basically characterized by a small diameter and a large power 
sign ratio. In addition, networks can be made \emph{typically} 
more resilient by adding more links.
\end{abstract}

\submitto{\NJP}
\maketitle

\section{Introduction}
Stability and control of power grids have not only been investigated by 
engineers \cite{Anderson03,Kundur94}, where rotor angle and voltage 
stability play an important role, but attracted also attention in the 
physics community. By generalizing the \textit{swing equation} 
\cite{Anderson03,Kundur94} of a synchronous machine to small networks by 
using the well studied Kuramoto model \cite{Kuramoto84,Kuramoto75},
Filatrella, Nielsen and Pedersen \cite{Filatrella08} stimulated many 
studies in this field. In \cite{Rohden12} the synchronization of this 
dynamical model on the topology of the power grid of the United Kingdom 
(UK) is investigated. While \cite{Magnesius12,Dorfler12} analyzed this 
Kuramoto-like model with respect to its application for power grids, 
in \cite{Brede08,Kelly11} synchrony optimized networks with 
Kuramoto oscillators were constructed. 
More generally, synchronization of oscillators 
with spectral methods was put under scrutiny in 
\cite{Barahona02,Donetti05,Motter13}.

In addition, many studies 
\cite{Sachtjen00,Motter02,Crucitti04,Crucitti04_2,Chassin05,Kinney05,Rosas_Casals07,Simonsen08,Sole08} 
deal with load models and analyze the vulnerability of networks due to 
node failures and the resulting cascading failures that might occur.

Here, we are not interested in the stability of dynamical systems but 
the resilience of networks. In \cite{Hartmann14} the resilience of 
a very basic model for transportation networks was investigated
via introducing a ``backup capacity'' (see below) and using
large-deviation techniques. The model in \cite{Hartmann14} assumes
that one unit of some ``quantity'' is transported between all
pairs of nodes along shortest paths. In the present work, we study
a very different model, still quite simple but  designed for modelling
power grids, based on the laws of electricity. Specifically, we
introduce a power flow model on networks based on the 
fixed points of a Kuramoto-like model \cite{Filatrella08,Rohden12} and a 
linearized DC power flow model (see e.g., \cite{Stott09}) as used in 
electrical engineering, respectively.

The resilience is defined by the 
\textit{backup capacity}. This quantity measures the overcapacity of the 
transmission lines which is needed to ensure stable operation when the 
most loaded link in the network exhibits a failure. We obtain the 
probability density functions (pdfs) of the resilience 
for three different random 
network ensembles and the topology of the power grid of the UK.

We are interested in obtaining the pdfs of these ensembles over 
a large range of the support, because a 
probability distribution contains the full information of a stochastic 
system in contrast to a finite number of moments 
(like e.g., mean or variance). To make statements about the 
different ensembles concerning the resilience we therefore need the 
whole pdf including the low-probability tails. By obtaining these tails 
we are also able to get properties of very vulnerable (high backup 
capacity) and very resilient (low backup capacity) networks. 
The analysis of these very resilient networks allows us to derive 
design principles for a resilient future power grid. 

Furthermore, given the backup capacity of an existing network, one 
can compare with a suitable network ensemble. The cumulative 
probability of finding a more resilient network (with smaller backup 
capacity) in the ensemble 
yields a quality measure for the investigated network. This is the 
so-called \textit{$p$-value}, which is a standard quantity in 
statistics to estimate the significance of a result. Sometimes, one 
needs to access the low probability tails of a pdf like in the present 
approach, where we want to study optimized high-resilience power grids.
This is 
analogous to the calculation of significance of protein alignments, 
where one also needs to access the tails of the pdf, since proteins are 
optimized by evolution \cite{Hartmann02}.
For an example of the $p$-value calculation, see \sref{pdf_PB}.

The paper is organized as follows. In \sref{model} the studied model 
and its simplification to static flow equations is described. 
\Sref{resilience} deals with the determination of the 
\textit{backup capacity} and hence the resilience of a network against 
transmission line failures. Next, \sref{networks} presents the 
investigated random network ensembles as well as the topology of an 
existing power grid. After this, the simulation and reweighting 
techniques are explained in \sref{LD_samp}, followed by \sref{results} 
which provides the numerical results of the simulations. Last, a 
conclusion is drawn and a short outlook is given.

\section{Model}
\label{model}
\subsection{Kuramoto-like model}
We use a simplified model of interconnected synchronous machines, 
derived from the dynamics of the rotor to model a power grid. The 
classic constant-voltage behind the transient-reactance model 
then gives 
the swing equations (see e.g., \cite{Anderson03,Dorfler12}) which are 
derived from energy conservation. The Kuramoto-like model used in 
\cite{Filatrella08,Rohden12,Dorfler12} is directly related to 
this swing equation. On each node $i$ of the network either a 
synchronous generator or a synchronous motor are placed. The generators 
exhibit power plants and therefore produce power 
($P_i^{\rm source} > 0$), whereas the motors consume power 
($P_i^{\rm source} < 0$).

Note that only active power is considered here and 
all transmission lines are regarded as lossless. 
These simplification appear justified to us, since for the
first time the resilience properties of models for electric power
grids are studied over the almost complete ensemble, even in the regime of 
extreme networks. Thus, our work lays a solid ground for later comparisons
to more sophisticated models. E.g., in 
\cite{Dorfler12,Anderson03} an 
extension of the model studied here to transmission lines with losses 
(using the admittance matrix) as 
well as nodal voltages is explained. A further expansion of the model 
with reactive power is given in \cite{Magnesius12}.

Each synchronous machine $i$ 
can be described \cite{Filatrella08,Rohden12} by its mechanical phase 
$\theta_i = \Omega t + \phi_i$, 
where $\Omega$ is the angular frequency ($2 \pi \cdot 50$ or 
$2 \pi \cdot 60$ rad s$^{-1}$) and $\phi_i$ the phase 
deviation. Note that the mechanical phase deviation $\phi_i$ is the same 
as the electrical angle $\delta_\rme$ except for a constant 
factor, namely the number of poles $n_{\rm{P}}$ of the synchronous 
machine \cite{Anderson03}: $\delta_\rme = (n_{\rm{P}}/2) \phi_i$.
To derive the equation of motion for $\phi_i$, one needs to consider 
energy (or power) conservation \cite{Filatrella08,Rohden12}, so that for 
each synchronous machine $i$
\begin{equation}
  P_i^{\rm source} = P_i^{\rm diss} + P_i^{\rm acc} + P_i^{\rm flow}, 
  \label{energy_cons}
\end{equation}
where $P_i^{\rm{source}} \gtrless 0$ depending on whether the machine is 
a generator or a motor. $P_i^{\rm diss}$ and $P_i^{\rm acc}$ are the 
dissipated and accumulated power, respectively. The power flow between 
two units $i$ and $j$ is given by \cite{Filatrella08,Rohden12}
\begin{equation}
	P_{ij}^{\rm flow} = -P_{ij}^{\rm MAX} \sin(\theta_j - \theta_i),
	\label{flow_P}
\end{equation}
where $P_{ij}^{\rm MAX}$ is the maximum capacity of the power line, 
which connects the nodes $i$ and $j$. The power flow of node $i$ is 
therefore given by the sum of the flow to all its neighbors
\begin{equation}
  P_i^{\rm flow} = -\sum_j P_{ji}^{\rm MAX} \sin(\phi_j - \phi_i),
\end{equation}
where we have used that $\theta_j - \theta_i = \phi_j - \phi_i$ and 
$P_{ij}^{\rm MAX} = 0$ if the edge between $i$ and $j$ does not exist.

From \eref{energy_cons} follows the equation of motion (for details 
see \cite{Rohden12,Filatrella08}) for the phase deviation of unit $i$
\begin{equation}
	\ddot{\phi}_i = P_i - \kappa \dot{\phi}_i + 
	\sum_j K_{ji}^{\rm MAX} \sin(\phi_j - \phi_i),
	\label{phase_dev}
\end{equation}
where $\kappa$ is a damping parameter. We apply uniform links,
i.e., $K_{ij}^{\rm MAX} = K^{\rm MAX}$ 
if a link exists between nodes $i$ and $j$ and $K_{ij}^{\rm MAX} = 0$ 
otherwise. The powers $P_i$ are directly related to $P_i^{\rm source}$ 
(cf., \cite{Rohden12,Filatrella08}). Note that \eref{phase_dev} is a 
version of the famous Kuramoto model \cite{Kuramoto75,Kuramoto84}.

\subsection{Simplifications leading to power flow model}
Here, we use a static approach, so we are only interested in the fixed 
points of \eref{phase_dev}. Hence, setting 
$\ddot{\phi}_i = \dot{\phi}_i  = 0$ yields
\begin{equation}
	P_i = -\sum_j K_{ji}^{\rm MAX} \sin(\phi_j - \phi_i).
	\label{static_eq}
\end{equation}

As we are not interested in the region close to the phase transition 
(where global synchronization sets in), as e.g., in 
\cite{Rohden12,Moreno04}, we choose a quite large value for the maximum 
capacity of the power lines. In fact, we choose $K^{\rm MAX} = 5 \cdot N$. 
Therefore, the argument of the sine in \eref{static_eq} needs to be small 
to fulfill the equation, as we choose the consumed and produced power 
$P_i$ uniformly from the interval $[-1,1]$, respectively. Hence, we 
expand the sine and obtain
\begin{equation}
	P_i = -\sum_j K_{ji}^{\rm MAX} (\phi_j - \phi_i),
	\label{static_lin_eq}
\end{equation}
which is a linear equation. It is independent of initial conditions and 
represents the power flow balance for each machine $i$. It is equivalent 
to the linearized DC (LDC) power flow model (for an overview see 
\cite{Stott09}) used in electrical engineering. The most popular variant 
of the LDC model uses the following simplifications \cite{Coffrin14} to 
derive the equations from the AC model. Please note that all 
approximations mentioned here also apply to the model studied in this 
work. First, the (absolute value of the) conductance of the transmission 
lines needs to be small in comparison with the susceptance 
(i.e., lossless lines corresponding to zero resistance). Second, the 
phase angle difference is small so that 
$\sin(\phi_j - \phi_i) \approx \phi_j - \phi_i$ holds. Third, the nodal 
voltages are $|E_i| \approx 1$ and constant over time.

\section{Resilience}
\label{resilience}
The observable to quantify the resilience of a network is based on the 
power flows between the synchronous machines in the power grid. These 
flows between two nodes $i$ and $j$ are basically given by \eref{flow_P}, 
where only the variables have been changed. Thus, we define
\begin{equation}
	K_{ij}^{\rm flow} = |K_{ij}^{\rm MAX} \sin(\theta_j - \theta_i)|,
	\label{flow_K}
\end{equation}
where we use here the absolute value to be independent of 
the direction of the flow. To calculate this power flow for all nodes 
the investigated network needs to be connected, i.e., no isolated nodes 
exist. In the sampling described in \sref{LD_samp} it is ensured 
that only connected networks are used.

To determine the power flows \eref{static_lin_eq} is solved numerically 
\cite{GSL} with given uniformly distributed $P_i \in [-1,1]$ and fixed 
$K^{\rm MAX} = 5 \cdot N$. The solutions for the phase deviations 
$\phi_i$ are then used in \eref{flow_K} to calculate the power flows for 
all links in the network. 

Next, the transmission line 
$e_{\rm max} = {\rm argmax}_{\{i,j\}} K_{ij}^{\rm flow}$ with the highest 
load (power flow) in the network is removed mimicking a failure in the 
transmission line. Selecting the highest-load line results
in a good estimate of the worst-case single-line failures 
\cite{Hartmann14}. 
Afterwards, \eref{static_lin_eq} is solved again, the 
power flows \eref{flow_K} of the network are recalculated resulting in 
flow values $\{\tilde{K}_{ij}^{\rm flow}\}$ of the modified network. Now, 
the \textit{backup capacity} is defined as the highest increase of the 
power flow over all edges
\begin{equation}
	P_{\rm B} = \max_{\{i,j\}} \; (\tilde{K}_{ij}^{\rm flow} - 
		K_{ij}^{\rm flow}).
\end{equation}

If the removal of $e_{\rm max}$ disconnects the network 
$P_{\rm B} = \infty$, so that these networks are neglected in the 
sampling. Due to the reorganization of the flow pattern, 
in some links also a decrease of the power 
flow is possible, so that some 
$\tilde{K}_{ij}^{\rm flow} < K_{ij}^{\rm flow}$. 

\begin{figure}[ht]
  \begin{center}
  \includegraphics[width=0.38\textwidth]{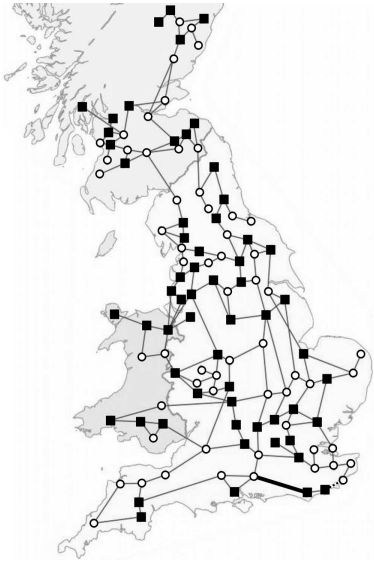}
  \end{center}

  \caption{Topology of the power grid of the United Kingdom (UK)
  \cite{UK_Grid_Deloitte,Rohden12,Simonsen08} with $N=120$ synchronous 
  machines and $M=165$ transmission lines. Generators ($P_i > 0$) are 
  labeled \opencircle and motors ($P_i < 0$) denoted by \fullsquare. Half 
  of the synchronous machines were chosen as generators and the other 
  half as motors. The dashed line in the south-east denotes the 
  transmission line $e_{\rm max}$ with the highest power flow 
  $|K^{\rm flow}| = 1.294$ which 
  connects machines with powers $P_{117} = -0.76$ and $P_{118} = 0.95$. 
  After the removal of this edge the power flow in the transmission line 
  depicted in bold (a bit to the west of the highest-flow line
  $e_{\rm max}$)
  increases most, actually it takes all the flow
  from $e_{\rm max}$ and it therefore defines the backup 
  capacity $P_{\rm B} = 1.294$. 
  Map of the UK from \cite{UK_map_url}, changed. 
  \label{UK_grid}}
\end{figure}

These backup capacities represent the resilience of a network against 
the failure of a transmission line. Note that the ability of a system 
to get back to stable operation after a single outage of a component 
(e.g., a transmission line) of a power grid is called $N-1$ criterion in 
electrical engineering. It is used in the planning and maintenance of 
power grids. Here, the backup capacities give an estimate of how much 
additional capacity of the transmission lines needs to be kept to keep 
them stable, even if one line breaks down. For small values of 
$P_{\rm B}$ the network structures are quite resilient, so few 
additional (over)capacity for the lines is needed. 
In case of large backup capacity 
the network structure does not allow to
compensate a single-line failure so easily, therefore 
it is less resilient.

\section{Networks}
\label{networks}
In this work we investigate one existing network and different network 
ensembles. We used the topology 
\cite{UK_Grid_Deloitte,Rohden12,Simonsen08} of the British 
power grid (see \fref{UK_grid}) to determine the resilience of this 
grid when generators and motors are randomly placed on the nodes.

In addition, we obtained the pdf by means of a histogram
of this resilience when one starts with the British grid 
with the same distribution of synchronous machines and 
uses the procedure for Erd\H os-R\'enyi graphs with fixed number 
of edges (cf., \sref{LD_samp}). A more detailed discussion about 
\fref{UK_grid} and the probability density function of 
the resilience is given in \sref{results}.

The studied network ensembles are the Erd\H os-R\'enyi (ER) graph 
ensemble \cite{Erdoes60}, 
and a spatial network ensemble \cite{Barthelemy11}. The different 
parameters in these network models are chosen such that each node has 
(on average) three neighbors. This should take into account that real 
transmission grids are sparse with an average number of neighbors per 
node of $\langle k \rangle \approx 3$. For the North American power 
grid Kinney et al.\ \cite{Kinney05} report about $N = 14\, 000$ substations 
and $M = 20\, 000$ transmission lines resulting in 
$\langle k \rangle \approx 2.9$. Watts and Strogatz \cite{Watts98}
found for the the electrical power grid of the western U.S.\ 
$\langle k \rangle = 2.67$.
For the European transmission grid, 
Sol\'e et al.\ \cite{Sole08} state a value of $\langle k \rangle = 2.70$.

The simplest type of random network is an ER random graph. In this ER
network ensemble \cite{Erdoes60} no assumptions on the topological 
structure of the network are made. It is therefore an ideal ensemble 
to be compared with, e.g., spatial networks to see the effects of 
topological structure.
The creation of an ER network works as follows. One starts with 
an empty network of $N$ 
nodes. Then, each pair $i$, $j$ of nodes is connected with the 
probability
\begin{equation}
	p_{ij}^{\rm ER} = c/N,
\end{equation}
thus $c = \langle k \rangle = 3$ is the connectivity of the network 
ensemble.

Next, we consider spatial networks \cite{Barthelemy11} which are embedded 
in a two-dimensional plane. Each of the $N$ nodes is distributed 
uniformly at random in a $[0,1] \times [0,1]$-plane, so to each node a 
$x$- and a $y$-position are assigned. A link is added between nodes $i$ 
and $j$ with probability
\begin{equation}
	p_{ij}^{\rm SN} = f \; \left( 1+ \sqrt{N \pi} \; d_{ij}/
				\alpha \right)^{-\alpha},
	\label{prob_spatial}
\end{equation}
where $d_{ij} = [(x_i-x_j)^2 + (y_i-y_j)^2]^{1/2}$ is the Euclidean 
distance between the two nodes. The parameters $f$ and $\alpha$ have 
been chosen such that an average number of neighbors 
$\langle k \rangle \approx 3$ is achieved. For all considered system 
sizes $N$ we used $\alpha = 3$ and $f \in [0.54,1.9]$ (in decreasing
order for increasing system size $N$).

In addition, we also used the ER ensemble with fixed number of links.

\section{Simulation and reweighting method}
\label{LD_samp}

To determine the pdfs over a large range of backup 
capacities for the different graph ensembles we use a reweighting 
technique. For details on the derivation of this technique we refer to 
\cite{Hartmann02,Hartmann11,Hartmann14} and state only the main ideas and 
results which are important for the determination of the pdf.

The main idea to reach very small probabilities or probability
densities of the order $10^{-100}$ is 
the use of an additional Boltzmann factor $\exp(-P_{\rm B}(G)/T)$ 
in a  
Markov-chain Monte Carlo (MC) simulation generating network instances.
This is different from \emph{simple sampling}, where the network
realizations are drawn directly and 
independently with their natural ensemble weights.
 The 
parameter $T$ is an artificial temperature, which makes it possible to 
sample different regions of the pdf of 
$P_{\rm B}$. The argument $G$ is the 
investigated network in the current MC step $t$. 

This MC simulation works as follows. In each step $t$ of the simulation 
a candidate network $G^*$ from the current network
$G(t)$ is created in the following way: First, a node $i$ is 
chosen uniformly at random. For the different network 
ensembles diverse techniques are now used. In the case of ER graphs all 
adjacent edges to $i$ are removed and with probability 
$p_{ij}^{\rm ER} = c/N$ a link is added for each other node $j$. 
For ER 
with a fixed number of edges also all adjacent edges to $i$ are removed. 
But next, node $i$ is connected with as many randomly chosen feasible 
nodes as removed edges. Hence, the number of edges is preserved.
For spatial networks 
the procedure is the same as for ER graphs, but the probability to add a 
link is now $p_{ij}^{\rm SN}$ (see \eref{prob_spatial}).

Next, it is checked whether the graph $G^*$ is connected. If this is not 
the case, the above procedure is repeated on $G$ until a feasible network 
$G^*$ is found. Note that also the initial networks need to be connected. 
Therefore, ring-type or complete (all $N (N-1)/2$ edges present) networks 
are created in the beginning and the MC simulations with the above 
described procedure run until the desired connectivity is reached.
For ER networks with fixed number of links the procedure for 
ER graphs with flexible number of links is used for this purpose.

After the candidate graph $G^*$ is created its backup capacity 
$P_{\rm B}$ is calculated. The candidate graph is then accepted 
($G(t+1) = G^*$) with the Metropolis probability
\begin{equation}
	p_{\rm Met} = \min \left\{ 1, \exp(-[P_{\rm B}(G^*) - 
	P_{\rm B}(G(t))]/T) \right\},
\end{equation}
otherwise the current graph is kept ($G(t+1) = G(t)$).

From \cite{Hartmann02,Hartmann11,Hartmann14}
\begin{equation}
	p(P_{\rm B}) = \exp(P_{\rm B}/T) Z(T) p_T (P_{\rm B})
	\label{Zet}
\end{equation}
one can determine the full pdf $p(P_{\rm B})$ with pdfs 
$p_T (P_{\rm B})$ measured at different finite temperatures $T$ up to a 
normalization constant $Z(T)$. This constant can be determined by 
choosing two histograms of neighboring temperatures. In the overlapping 
region the pdfs need to agree which makes it possible to 
calculate $Z(T)$ via \eref{Zet}. In an iterative procedure the histograms 
are ``glued'' together until the full pdf is obtained. A more 
detailed explanation with examples about the merging of the different 
histograms is given in \cite{Hartmann04}.

In order to check if the MC simulations are equilibrated, two different 
initial networks are used: A ring graph, where all nodes have two 
neighbors and a complete (fully connected) graph. Equilibration 
is reached when both values of $P_{\rm B}$ agree within the range of 
fluctuations. For the ensemble with fixed number of links, 
we studied the average of $P_{\rm B}$ over MC sweeps to determine the 
equilibration time. 
The longest equilibration time we observed was 
$2 \cdot 10^6$ sweeps for ER networks with $N=50$ nodes.

\section{Results}
\label{results}

We performed simulations for the UK grid, ER networks, spatial 
networks and ER graphs with a fixed number of links. For all networks 
except the UK grid we used networks of size $N=10$ up to $N=400$. For ER 
and spatial graphs the determination of the (possibly) full pdf 
of the resilience for $N=400$ was not possible, as a large gap in this 
pdfs appeared. The histograms for different temperatures have 
their peak either below or above this gap, which made it almost impossible 
to sample in this gap. One way to overcome this is Wang-Landau sampling 
\cite{Wang01}, which we did not try.
Nevertheless, we took data also for these two ensembles for $N=400$ to 
analyse other quantities (see \sref{res_vul_NWs}).

After leaving out the data before equilibration time and taking
samples only in intervals such that the Markov chain is roughly
decorrelated, our final data sets contain between 
about $5 \cdot 10^3$ samples for $N=400$ up to 
almost $10^7$ samples for $N=10$.

The consumed and produced powers $P_i$ are drawn from a uniform 
distribution $P_i \in [-1,1]$ and sum up to zero $\sum_i P_i = 0$. 
Furthermore, a combination of $\{P_i\}$ is chosen, such that the same 
number of generators ($P_i > 0$) and motors ($P_i < 0$) are drawn.
Most simply, sets of random numbers from the interval $[-1,1]$ 
were drawn until 
all above criteria can be fulfilled by assigning the last node.
This is a bit time consuming, 
but has to be performed only once during a simulation: The
power values
attached to the nodes 
are not changed during the Markov-chain MC since all possible
networks can be accessed just via changing the edges.

\subsection{Probability density functions of the backup capacity}
\label{pdf_PB}
First, we analyze the probability density functions of the backup 
capacity 
for the different network ensembles as well as for the UK power grid.

\begin{figure}[hb]
  \raggedleft
  \subfigure{\includegraphics[width=0.375\textwidth]
		{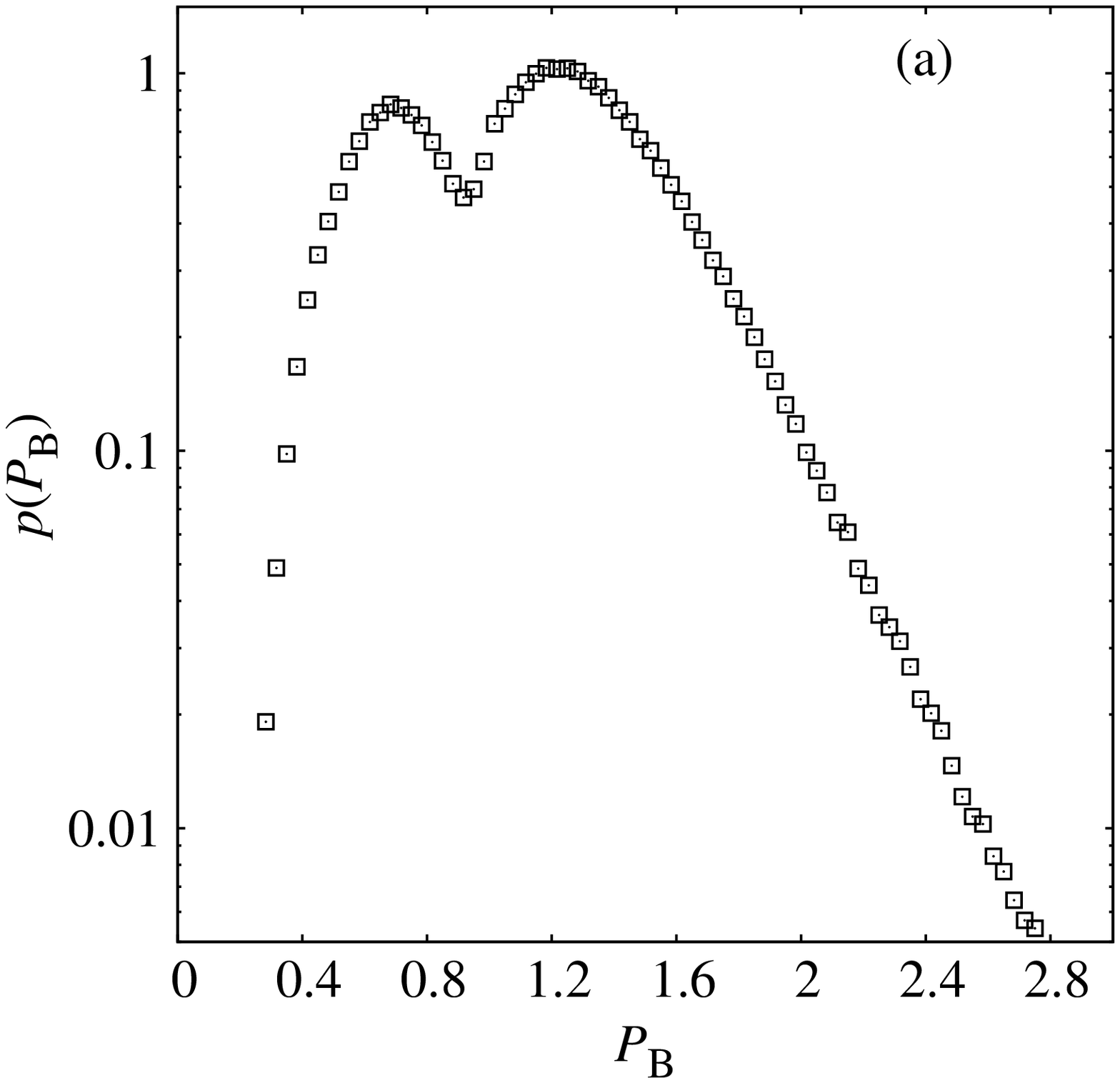}}
  \hspace{1mm}
  \subfigure{\includegraphics[width=0.4\textwidth]
		{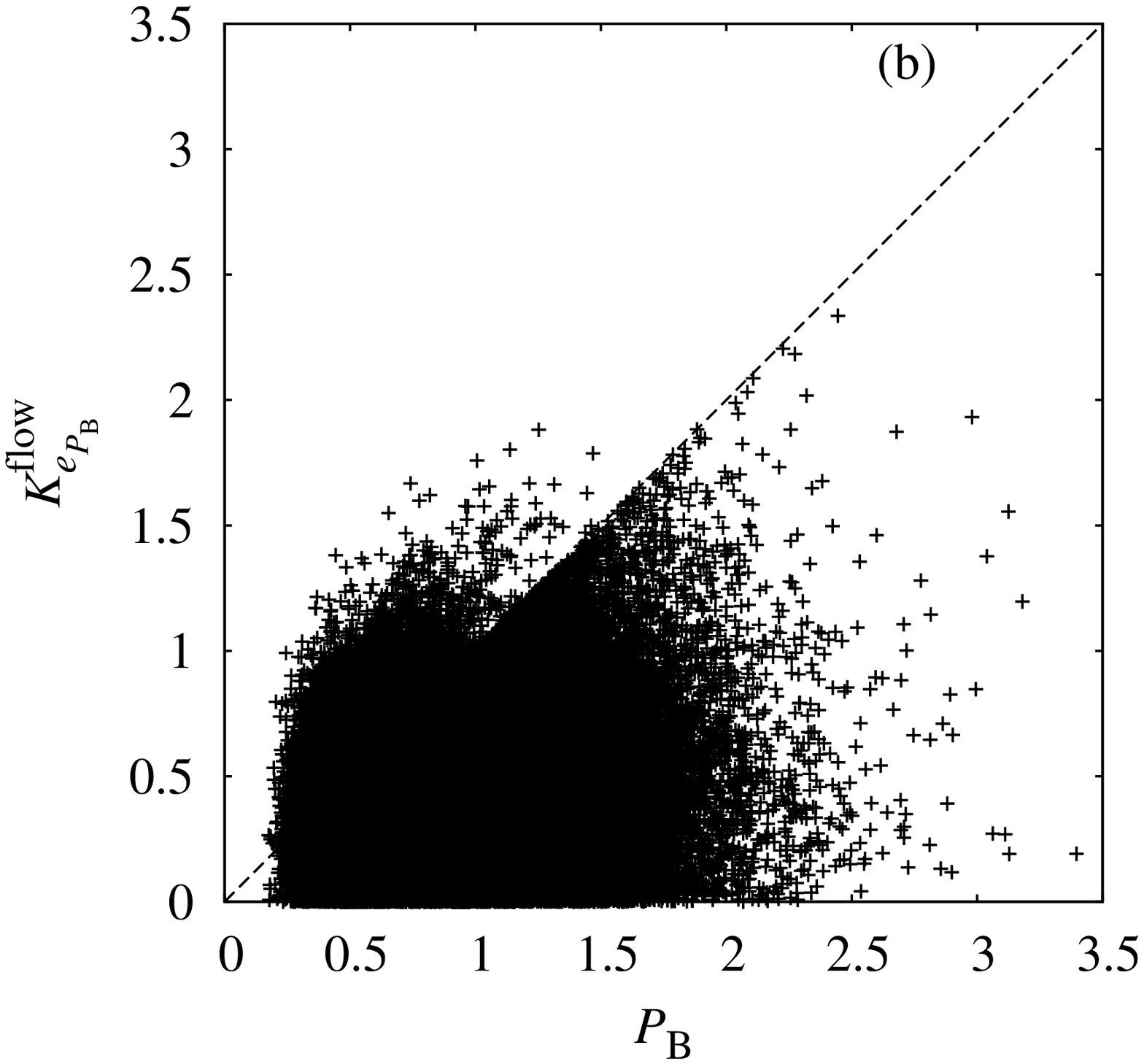}}

  \caption{{(a) Probability density function} $p(P_{\rm B})$ of the 
  backup capacity 
  $P_{\rm B}$ for ER networks with $N=120$ and $M=165$ (fixed) starting 
  from the UK grid (see \fref{UK_grid}). About $10^6$ samples are used to 
  generate this pdf. \newline
  {(b) Scatter plot of the power flow $K_{e_{P_{\rm B}}}^{\rm flow}$ 
  before the removal of the highest-load link through the edge which 
  exhibits the highest flow increase (i.e., later defines the backup 
  capacity) against the backup capacity $P_{\rm B}$ for $N=100$, an ER 
  ensemble and $10^5$ samples. Dashed line represents 
  $K_{e_{P_{\rm B}}}^{\rm flow} = P_{\rm B}$.} 
  \label{dis_UK_grid}}
\end{figure}

\Fref{dis_UK_grid}(a) shows the pdf of the resilience for ER 
networks with a fixed number of links and $N=120$ nodes based on the UK 
grid (cf., \fref{UK_grid}). The procedure to create a candidate graph in 
the large-deviation scheme is the same as for ER graphs with a fixed 
number of edges (see \sref{LD_samp}). Note that we only used one 
histogram with simple 
sampling corresponding to the temperature $T = \pm \infty$. Nevertheless, 
backup capacities from smaller than $0.4$ up to 2.8 could be measured.
In \fref{dis_UK_grid}(a) we can see 
an interesting double peak structure of the pdf, where the right 
peak is higher than the left. In these peaks the networks with typical 
values of the backup capacity are represented like the initial network 
of the simulation.
A closer look to the two peaks reveals:
Consider the power flow $K_{e_{P_{\rm B}}}^{\rm flow}$ before 
the removal of the highest-load link through the edge $e_{P_{\rm B}}$ which 
later defines 
the backup capacity, i.e., exhibits the highest flow increase. 
In \fref{dis_UK_grid}(b), where $K_{e_{P_{\rm B}}}^{\rm flow}$ is 
plotted against the backup capacity, 
two clusters become visible: One cluster is represented 
by a very small flow 
through $e_{P_{\rm B}}$ before removal of $e_{\rm max}$ and quite high 
values of $P_{\rm B}$ (cluster below the dashed line). 
This cluster corresponds to the right (higher) peak of the pdf. 
An explanation for the left peak is that it belongs to the 
cluster, where a considerable flow through $e_{P_{\rm B}}$ is already 
present in the network and thus, the flow increase $P_{\rm B}$ is rather 
small (cigar-shaped cluster at small $P_{\rm B}$, many points 
above the dashed line).

The UK grid (see \fref{UK_grid}) has a backup 
capacity of $P^*_{\rm B} = 1.294$ which is in this region of 
typical networks. Nevertheless, in principle much more
resilient networks exist.
This is confirmed by the $p$-value of the UK grid shown in 
\fref{UK_grid}. To obtain the $p$-value we calculate the cumulative 
probability that networks with smaller (or equal) backup capacity exist in 
the ensemble: $P(P_{\rm B} \leq P^*_{\rm B}) \approx 0.67$. This value 
tells us that the probability of finding a more resilient network than the 
UK grid in the ER ensemble with fixed edges ($N=120$, $M=165$) is higher 
than 2/3. This means the UK grid as depicted in \fref{UK_grid} has a low 
significance in terms of resilience, because of the large $p$-value.
The right tail of the pdf follows an 
exponential resulting in a line in a logarithmic plot, whereas the left 
tail is much more curved.

\begin{figure}[hb]
  \raggedleft
  \subfigure{\includegraphics[width=0.41\textwidth]
			{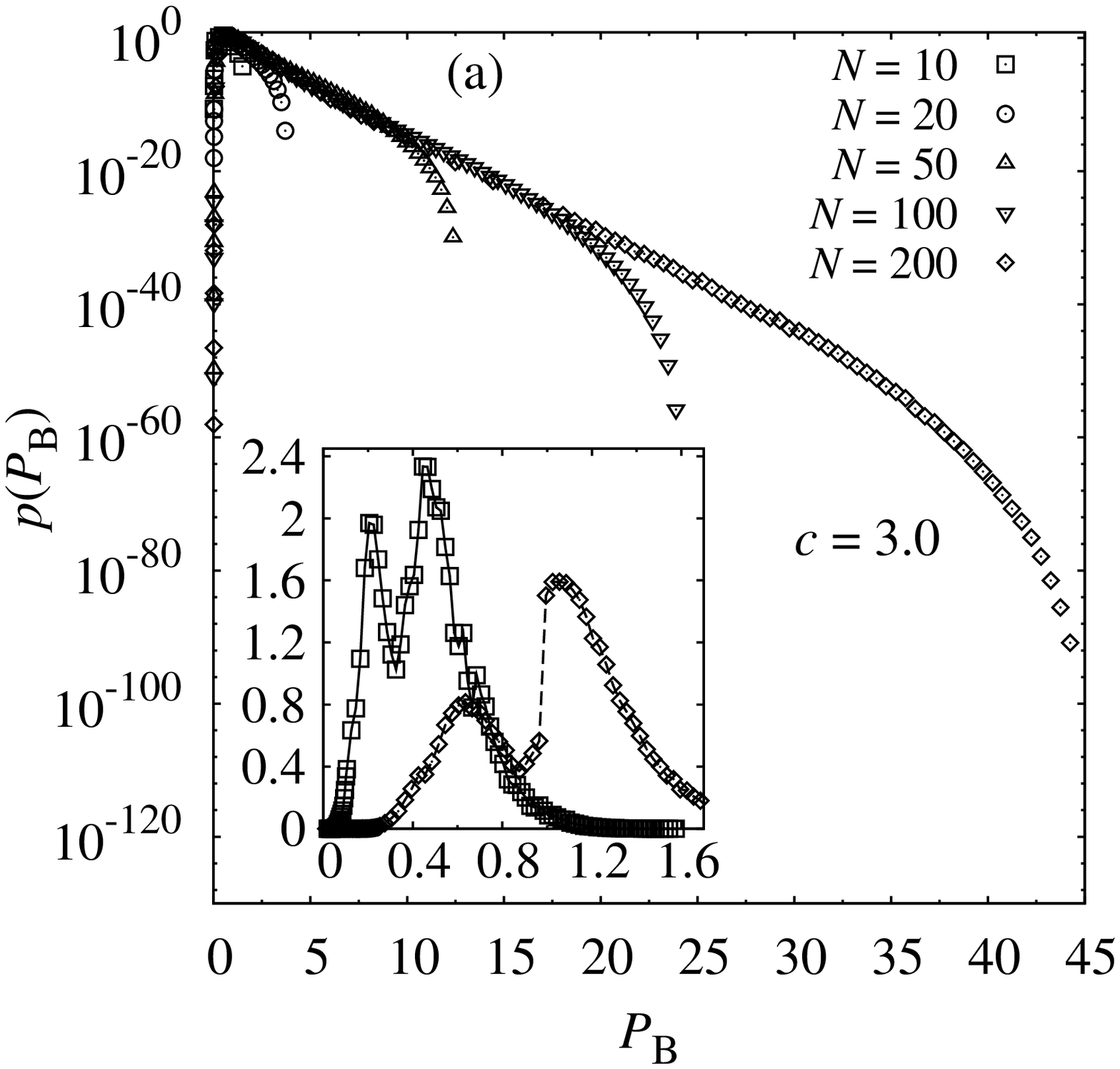}}
  \subfigure{\includegraphics[width=0.39\textwidth]
			{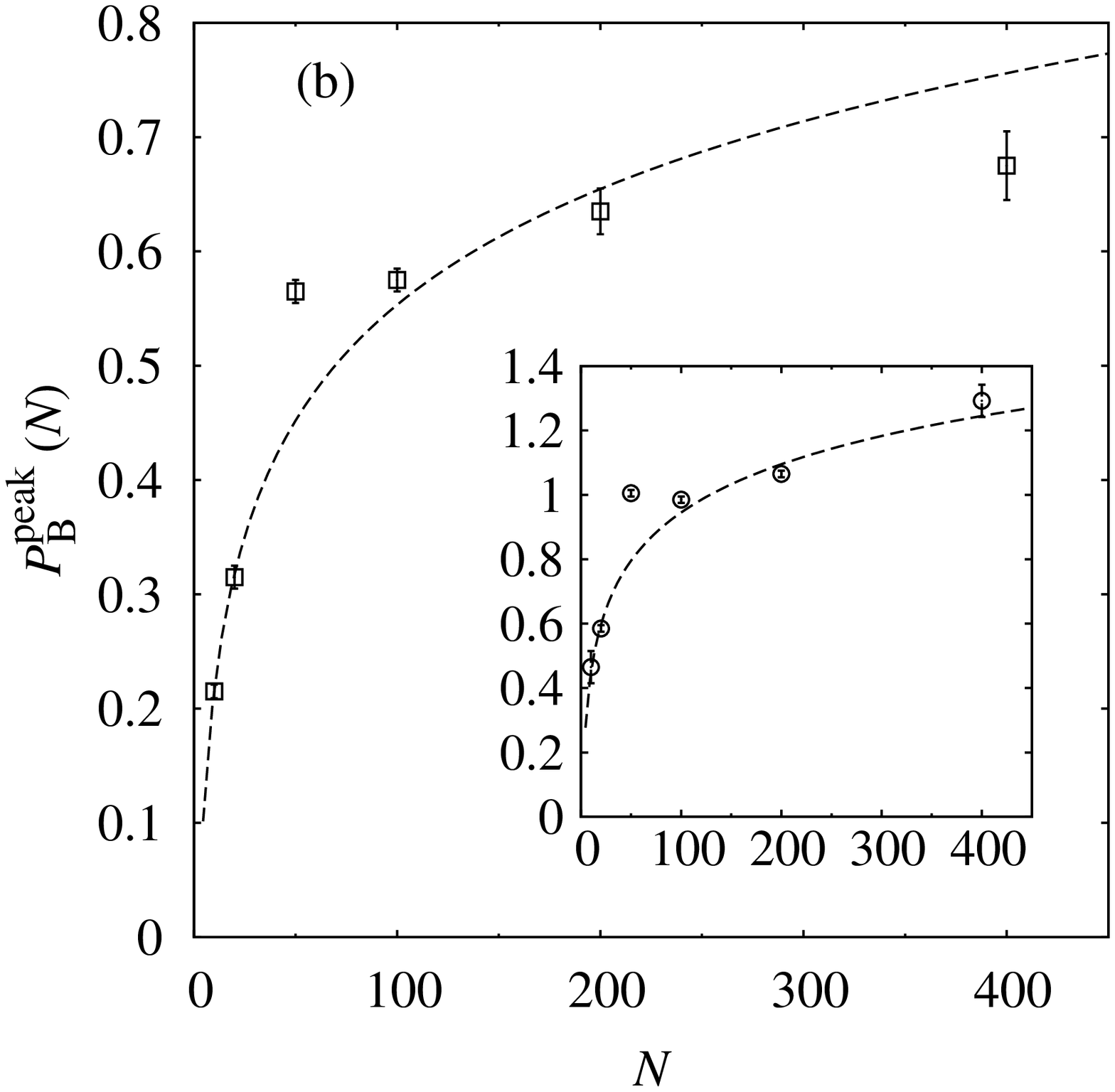}}
  \caption{(a) {Probability density functions} $p(P_{\rm B})$ of the 
  backup capacity 
  $P_{\rm B}$ for ER networks with sizes $N=10$ up to $N=200$.
  Inset: 
  Region close to the peaks of the pdf for $N=10$ and $N=200$. 
  Lines are guides to the eyes only. \newline
  (b) Peak positions $P_{\rm B}^{\rm peak}$ of the left peak in the 
  pdfs for ER networks as a function of network size $N$.
  Dashed line is a logarithmic fit ($N=50$ excluded)
  $P_{\rm B}^{\rm peak} (N) = a \cdot \ln(b \cdot N)$ with parameters 
  $a = 0.146(8)$ and $b = 0.44(6)$. 
  Inset: The same for the right peak. The
  dashed line is a logarithmic fit ($N=50$ excluded)
  $P_{\rm B}^{\rm peak} (N) = c \cdot \ln(d \cdot N)$ with parameters 
  $c = 0.216(17)$ and $d = 0.79(26)$.
  \label{ER_dis}}
\end{figure}

Next, we compare the results for ER networks, spatial networks, 
and ER graphs with a fixed number of links. For all these ensembles we 
obtained the pdf over the possibly 
full support of $P_{\rm B}$. 
For many of the pdfs it was quite difficult to obtain the very 
left or very right tail. For very small values of the backup capacity 
(corresponding to small positive temperatures in the large-deviation 
approach) the histograms tend to become delta shaped, meaning the 
observable $P_{\rm B}$ becomes almost constant over MC time. The large 
values of the backup capacity (i.e., small negative temperatures) are even 
more difficult to obtain. In the simulations one could see that the 
maximum value of $P_{\rm B}$ could hardly be reached in the pdfs 
as sampling in this region results in a delta distribution. This is also 
visible in the finally obtained pdf 
(cf., e.g., \fref{ER_dis}(a)), because as the values of 
$P_{\rm B}$ get closer to the maximum a strong curvature appears.

\Fref{ER_dis}(a) shows the pdfs of the backup capacity for ER 
networks on almost the full support for different graph sizes $N$. In the 
inset of \fref{ER_dis}(a) the double-peaked structure of the pdfs 
for the smallest and the largest obtained networks are shown. For 
increasing graph size $N$ the double peaks shift towards larger backup 
capacities. We found that this shift is logarithmic in $N$
(cf., \fref{ER_dis}(b)). 
Interestingly, the right peak becomes more pronounced for 
$N=200$ in comparison to $N=10$.

With the large deviation approach described in \sref{LD_samp} one is able 
to access typical, very resilient and very vulnerable networks. Typical 
networks close to the double peaks of the pdfs show a rather 
small backup capacity. In \fref{ER_dis}(a) very vulnerable networks with a 
large backup capacity of $P_{\rm B} \approx 45$ for $N=200$ are very rare 
and appear only with a probability density of about $10^{-90}$. Note that 
such small probabilities (densities)
are impossible to reach with ordinary MC simulations. These vulnerable 
networks are located at the right tail of the pdf. Although
this tail is compatible with an exponential a strong curvature 
occurs when 
the maximum possible value of the backup capacity is approached.
For transportation networks \cite{Hartmann14} the right tail of the 
pdf does not show any curvature.
The very resilient networks can be found in the left tail close to the 
peaks of the density function.

\begin{figure}[ht]
  \hspace{2.45cm}
  \subfigure[ER networks with fixed number of links and sizes $N=10$ 
  up to $N=400$. \label{ER_fixed_dis}]
  {\includegraphics[width=0.387\textwidth]
			{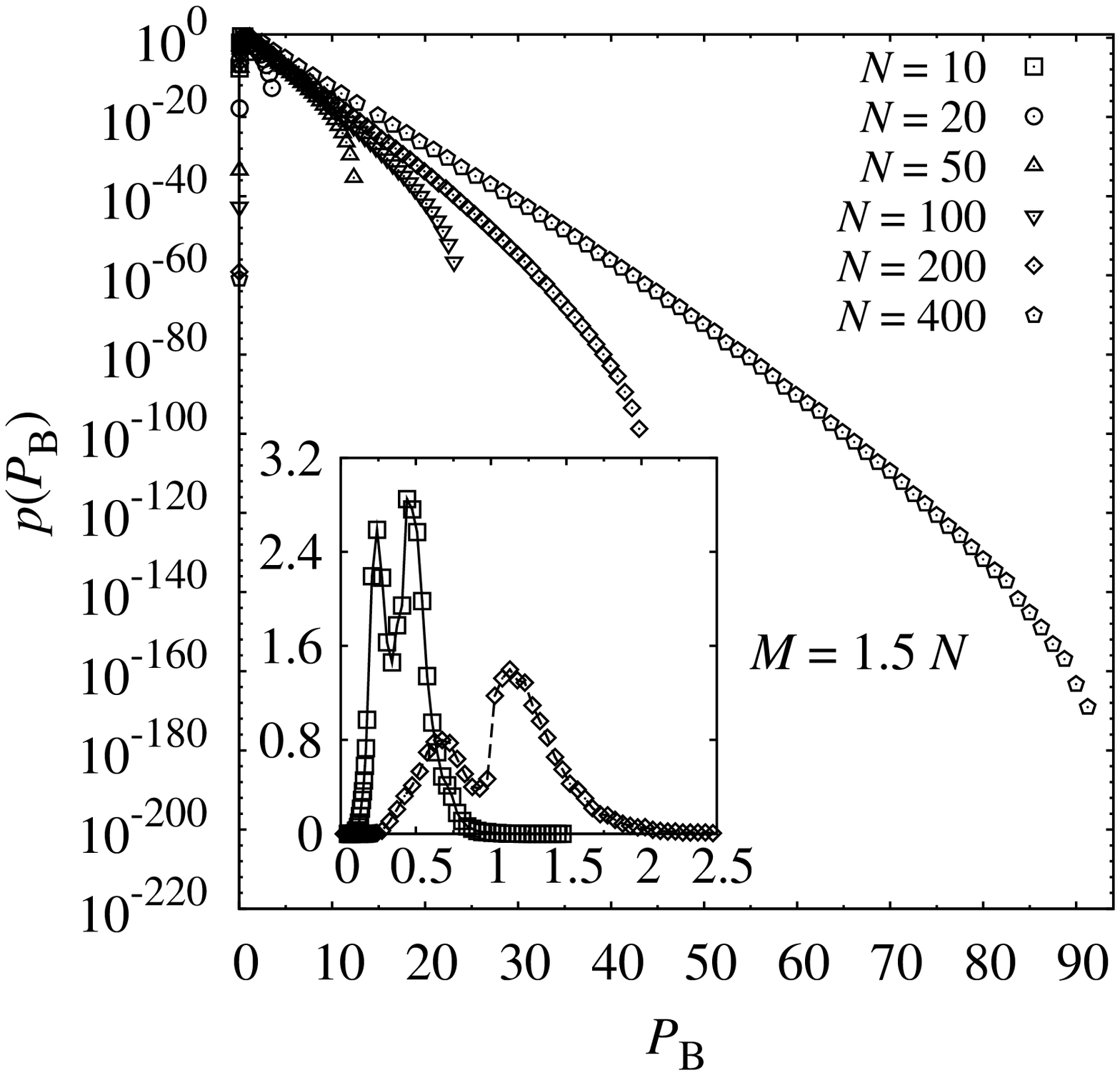}}
  \hspace{1mm}
  \subfigure[Spatial networks with sizes $N=10$ up to $N=200$.
  \label{SN_dis}]
  {\includegraphics[width=0.378\textwidth]
			{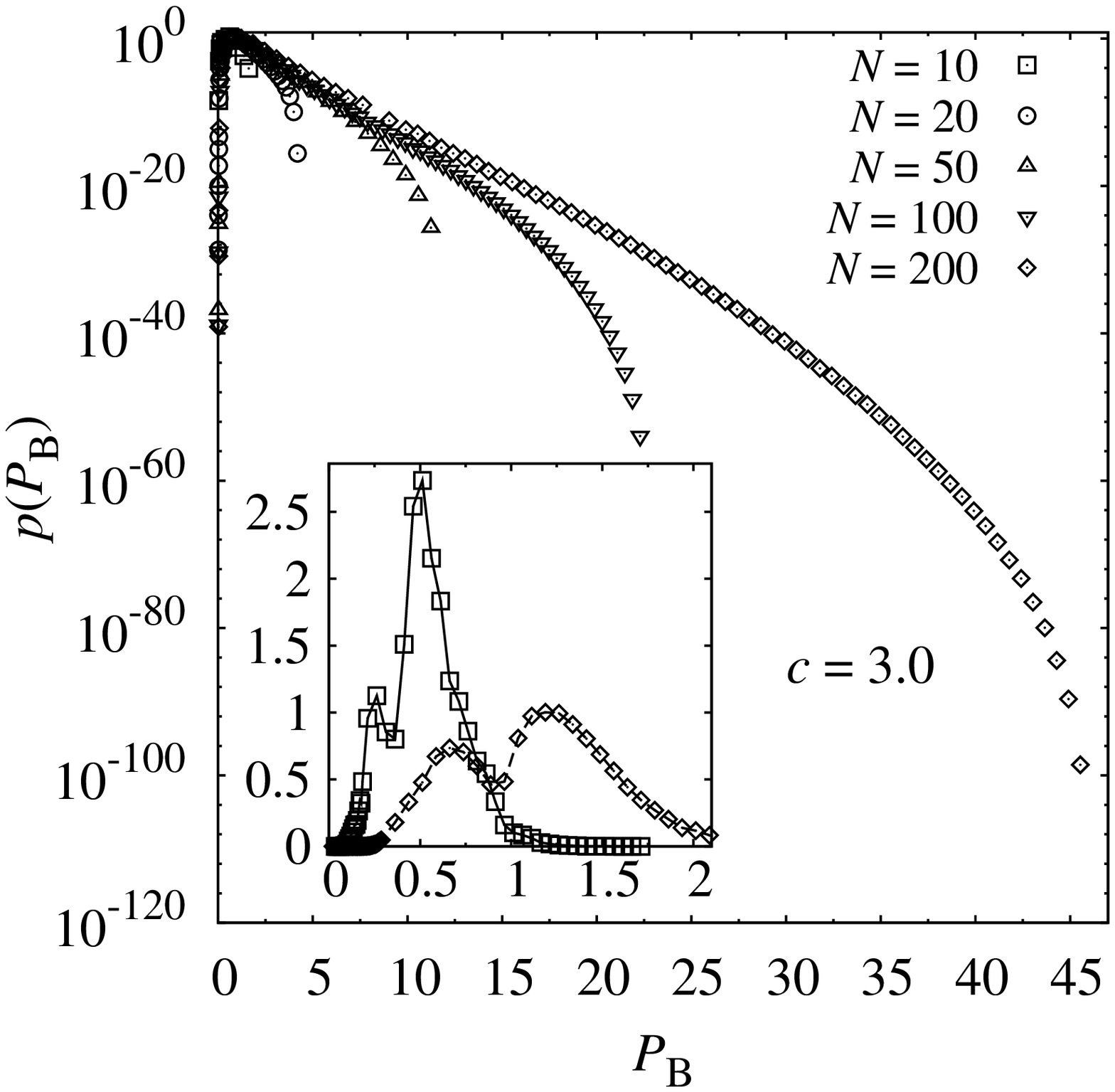}}
  \caption{Probability density functions $p(P_{\rm B})$ of the 
  backup capacity 
  $P_{\rm B}$ for given networks. Inset: Region close to the peaks of the 
  pdf for $N=10$ and $N=200$. Lines are guides to the eyes only.}
\end{figure}

In \fref{ER_fixed_dis} the pdf for 
ER graphs with fixed number 
of edges is shown. Again, the peaks move logarithmically to the 
right with increasing $N$. Like for ER networks with variable number 
of links, the right of the two peaks is almost twice as high as the left 
peak for $N=200$. In contrast, for $N=10$ the peaks have almost the same 
height.
The curvature of the right tail is not as strong as for ER or 
spatial networks.

\Fref{SN_dis} shows the results for the spatial network model, where the 
nodes of the network are placed in a two-dimensional plane. As found for 
the other network ensembles the peaks in the inset of \fref{SN_dis} shift 
logarithmically
towards the right with growing number of nodes. The 
two peaks differ less in height for $N=200$ than for $N=10$. The curvature 
of the right tail is stronger than for the ER ensemble with fixed 
number of links.

Next, we compare the resiliences of typical, very vulnerable and very 
resilient networks for the different ensembles. For the typical networks 
we investigate the right (usually higher) 
peak of the pdf for $N=200$. For 
the ER network ensemble 
the typical networks exhibit a quite small backup capacity 
of $P_{\rm B} \approx 1.065$, followed by the ER ensemble 
with fixed number of links ($P_{\rm B} \approx 1.115$). 
Almost as resilient are the 
typical spatial networks with $P_{\rm B} \approx 1.163$. 
Similar results are found for the left peak.

Very vulnerable networks at $P_{\rm B} \approx 44$ for $N=200$ are 
most unlikely for the ER ensemble with fixed number of links 
($p(P_{\rm B} \approx 44) \approx 10^{-106}$). For the ER 
ensemble this 
probability density at $P_{\rm B} \approx 44$ is about $10^{-88}$. 

Networks from the spatial network ensemble 
have densities of about $10^{-82}$ at 
$P_{\rm B} \approx 44$. These results support the 
findings for the typical networks as for the two ER ensembles it is 
unlikely to find very vulnerable networks, i.e., with a large backup 
capacity. The graphs from the spatial network ensemble exhibit 
the highest densities at 
large backup capacities and thus favor less resilient networks.

Very resilient networks at $P_{\rm B} \approx 0.024$ for $N=200$ are most 
unlikely for the ER network ensemble with a fixed number of links, 
where $p(P_{\rm B} \approx 0.024) \approx 10^{-52}$. 
For the spatial ($p(P_{\rm B} \approx 0.024) \approx 10^{-37}$) and 
ER ($p(P_{\rm B} \approx 0.024) \approx 10^{-34}$) network ensemble 
the densities to find a network with $P_{\rm B} \approx 0.024$ are almost 
equal.
In contradiction to what 
has been found previously, the ER ensemble with a fixed number of 
links exhibits 
quite low densities around $P_{\rm B} \approx 0.024$. Both spatial and ER 
network ensembles favor very resilient graphs, which have a small 
backup capacity.

These results need to be taken with care as for the ensemble with 
fixed number of edges arbitrarily small 
backup capacities can not be reached. In contrast, for 
the two ensembles with a flexible number of links many edges are 
allowed to be present in the networks and especially, 
the complete graph (each node is connected to all other nodes) is included 
in these ensembles. Therefore, the probability densities at 
low $P_{\rm B}$ 
for the ensembles with flexible number of links are of order $10^{18}$ 
higher than for the ER ensemble with fixed number of links.

To sum up, the most promising network ensemble in terms of resilience is 
the ER ensemble, although it is high dimensional, i.e., quite unrealistic. 
The ER ensemble with fixed number of links is also quite resilient 
against transmission line failure. The more realistic (it is embedded in a
two-dimensional plane) spatial network ensemble is also a good candidate 
for choosing resilient networks, although vulnerable networks are 
quite likely. These findings are compatible with \cite{Hartmann14}, 
although a much simpler very general transport model was studied there.

\begin{figure}[h]
    \hspace{2.5cm}
    \subfigure{\includegraphics[width=0.371\textwidth]
    {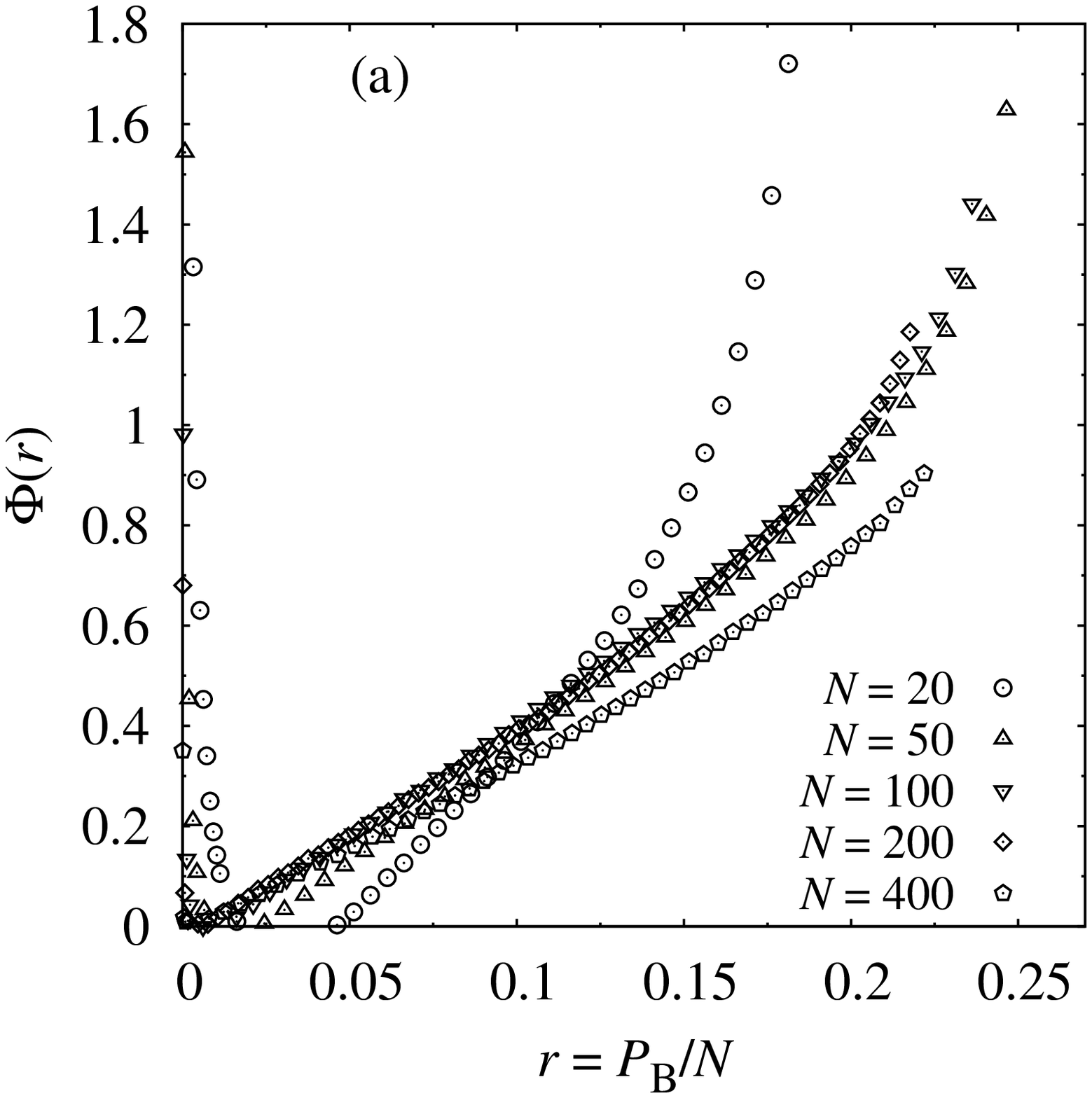}}
    \hspace{1mm}
    \subfigure{\includegraphics[width=0.4\textwidth]
    {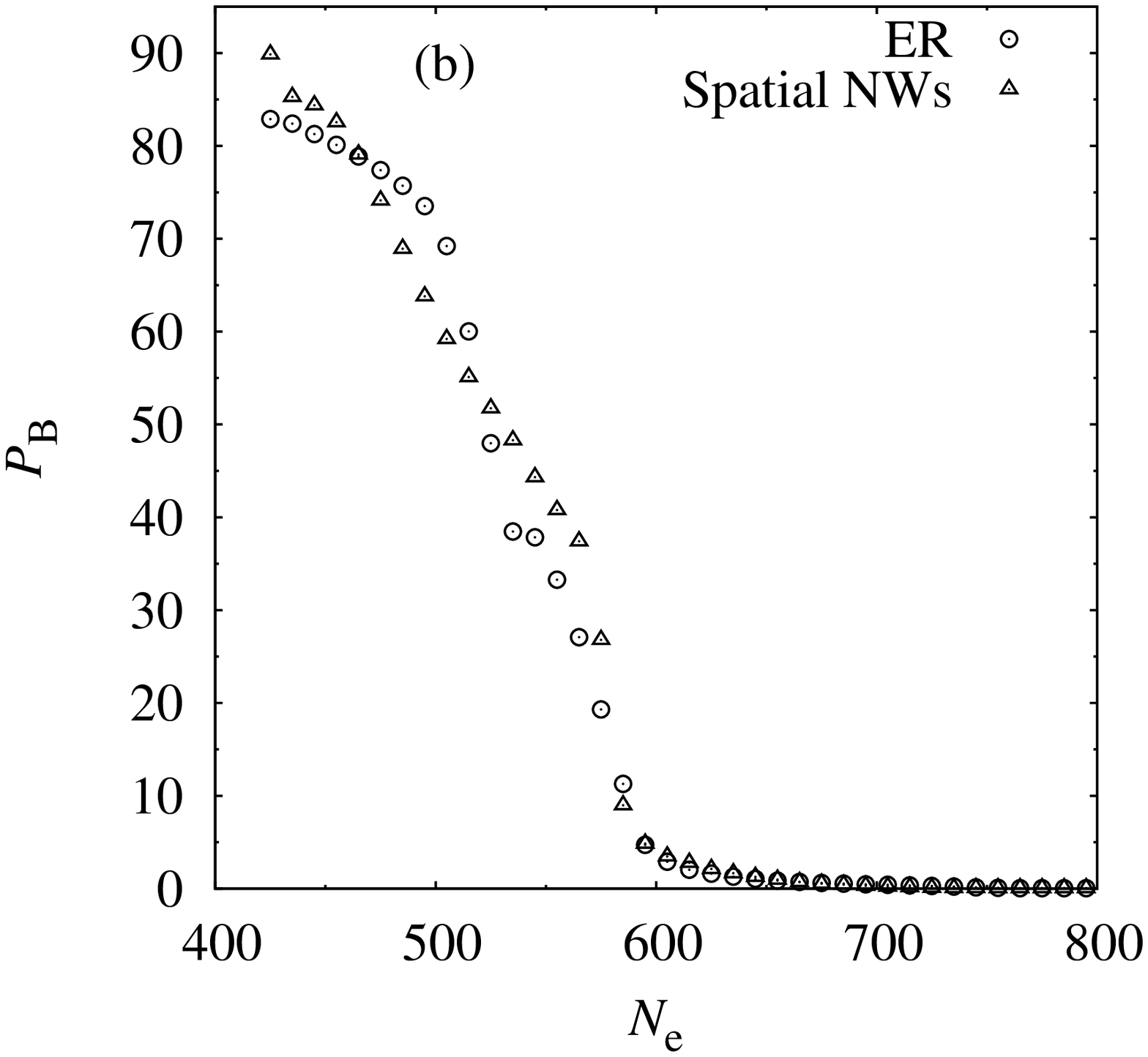}}
    \caption{(a) Rate function $\Phi$ as a function of the rescaled backup 
    capacity $r = P_{\rm B}/N$ for ER networks with fixed number of edges 
    ($M = 1.5 N$) and different sizes $N$. \newline
    (b) Average backup capacity $P_{\rm B}$ as a function of the number 
    of edges $N_{\rm e}$ for ER and spatial networks with size $N=400$.
    \label{rate_fct_PB_num_edges}}
\end{figure}

\subsection{The rate function}
Next, we investigate the behavior of the so called \textit{rate function}
\cite{Hollander00,Touchette09}
\begin{equation}
	\Phi = -\frac 1 N \log p(P_{\rm B}),
\end{equation}
which describes that the leading behavior (away from the typical 
instances) of the pdf is an 
exponential decay $p(P_{\rm B}) \sim e^{-N P_{\rm B}}$. 
\Fref{rate_fct_PB_num_edges}(a)
shows the rate function as a function of the rescaled backup capacity 
$r = P_{\rm B}/N$ for {the ER ensemble} with fixed number of links. 
This rescaling is motivated by the 
following observation. Consider a network which consists of two large 
$\mathcal{O}(N)$ subnetworks which are connected via a core (e.g., a 
triangle, cf., \cite{Hartmann14}) with generators on one side and motors 
on the other. In this setup the most power has to flow through a single 
link of the core. After the removal of this high-loaded link, the power 
flows through the other two core links of the triangle. Hence, the backup 
capacity increases by an amount of $\sim N$, as the power flows of $N/2$ 
nodes run through these two links.

In \fref{rate_fct_PB_num_edges}(a) one sees that the rate function 
approaches a limiting 
curve as $N$ increases. Below a certain value $r = r^*$ this curve is 
approached from below, whereas above $r^*$ the curve is approached from 
larger values of $\Phi$. The point $r^*$ moves towards smaller 
values of $r$ as $N$ increases. 
For $N=400$ deviations from this behavior for large $r$ can be 
observed.
Although the limiting curve is not compatible with a straight 
line as in \cite{Hartmann14}, still an exponential 
behavior with strong curvature for $p(P_{\rm B})$ is possible.

Note that the rate functions for the other graph ensembles look similar 
with basically the same limiting behavior.
This apparent convergence of the empirical rate function 
indicates that it might be
promising to apply analytical 
large-deviation techniques \cite{Hollander00,Touchette09}
to study resilience of power grids for these graph ensembles.

\subsection{Characterization of very resilient and very vulnerable networks}
\label{res_vul_NWs}

Next, we investigate the relationship between the backup capacity, i.e., 
resilience and the number of edges in the graph for the ER and 
spatial network ensemble. Hence, we used our simulation results to bin 
data jointly for all different 
temperatures $T$ with respect to the number $N_{\rm e}$ of edges. In each 
of these bins the average backup capacity is calculated and the result is 
shown in \fref{rate_fct_PB_num_edges}(b).
For a small number of edges in the network the backup capacity is very 
large. However, for many edges $P_{\rm B}$ assumes very small values. This 
means that in general a network with more edges is more resilient than a 
network with fewer edges. Note that adding a link to a network can 
sometimes destabilize it according to Braess's paradox \cite{Witthaut12}. 
With our data it is not possible do determine whether the steep decrease 
of the backup capacity appears at smaller $N_{\rm e}$ for the ER or 
spatial network ensemble. In \cite{Hartmann14} the decrease appears at a 
smaller number of 
edges for the ER ensemble. Thus, in contrast to general 
transportation networks, it is possible to obtain very resilient
power grids embedded in a two-dimensional plane with the same effort, i.e.,
number of edges, as for an infinite-dimensional, i.e., less
restricted, ER ensemble.

\begin{figure}[h]
  \hspace{2.48cm}
  \subfigure[Average diameter $d$. Inset enlarges region for small 
  $P_{\rm B}$. \label{diameter}]
  {\includegraphics[width=0.375\textwidth]
				{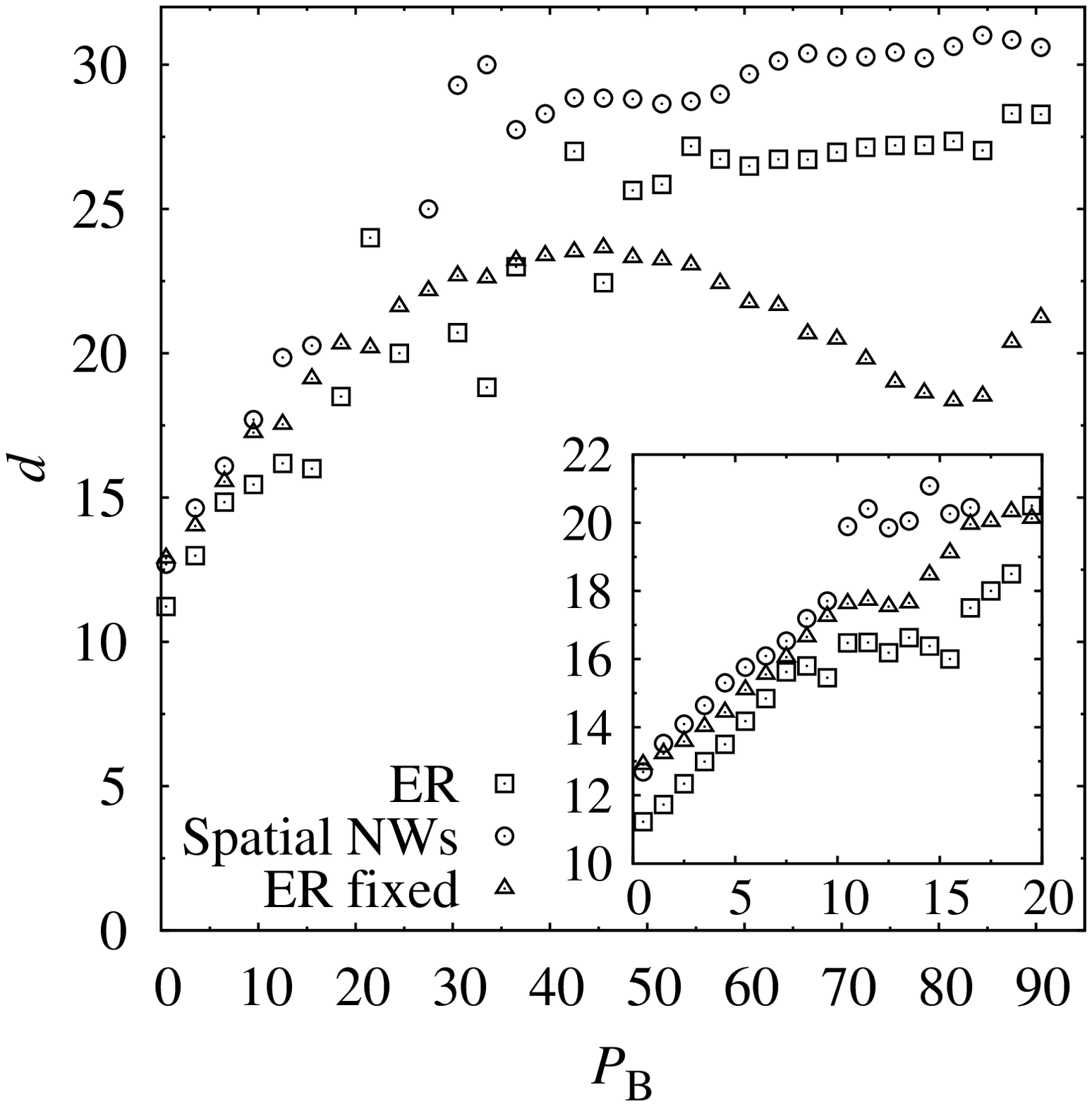}}
  \hspace{1mm}
  \subfigure[Average power sign ratio $p_{-}$. \label{power_sign_ratio}]
  {\includegraphics[width=0.384\textwidth]
				{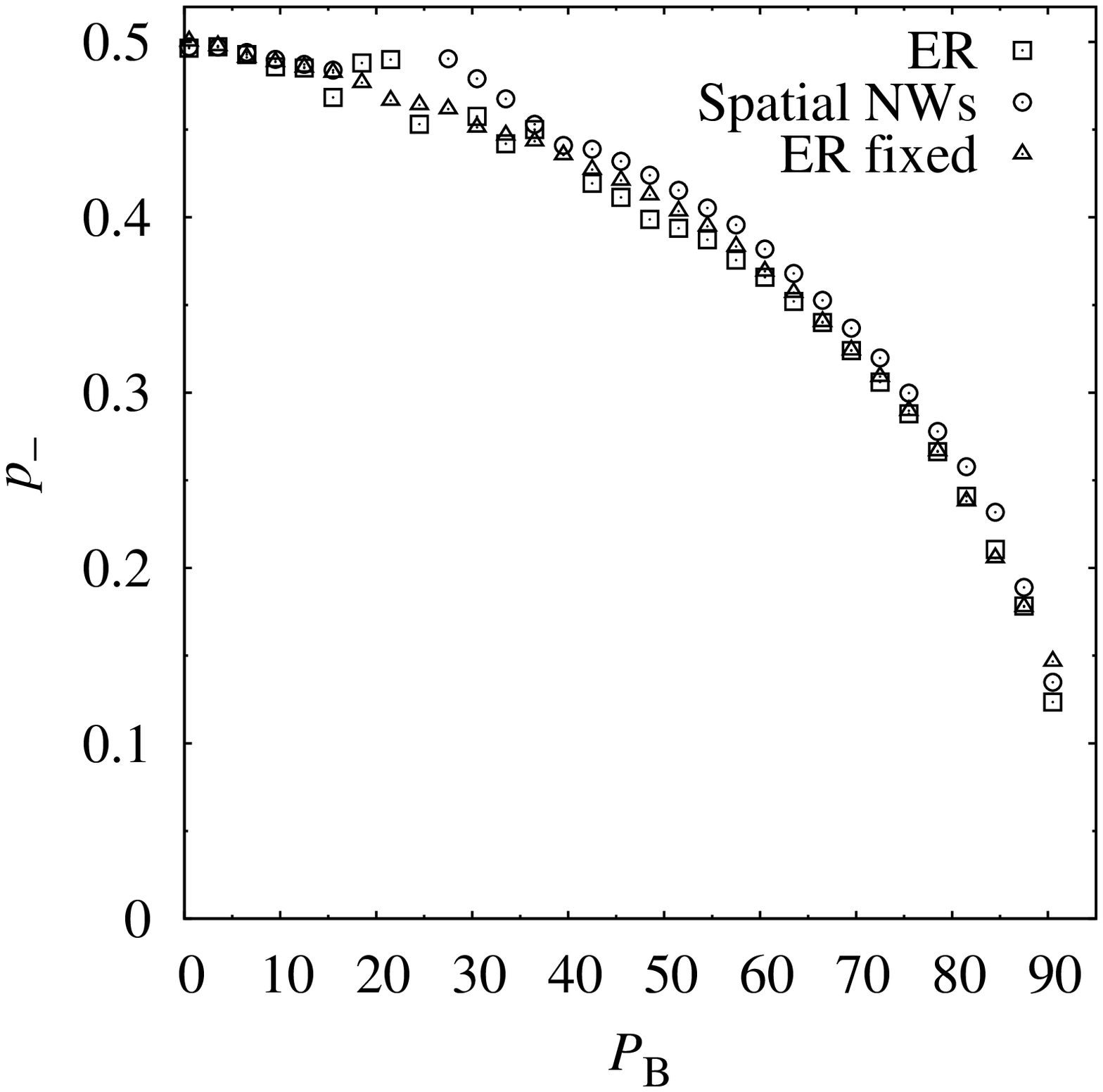}}
  \caption{Average diameter $d$ and power sign ratio $p_{-}$ as a function 
  of the backup capacity $P_{\rm B}$ for all studied network ensembles 
  with size $N=400$. 
  Note that because of the gap in the data for ER and 
  spatial networks some intermediate values of $P_{\rm B}$ are not 
  present.}
\end{figure}

\Fref{diameter} shows the average diameter for all studied network 
ensembles and $N=400$. The diameter of a network is defined as the longest 
of the shortest paths between all possible node pairs. First, a binning of 
the data with respect to the backup capacity is performed and afterwards 
the diameters are averaged within each bin. In the inset of 
\fref{diameter} one can see that with increasing backup capacity the 
diameter also increases at least for small values of $P_{\rm B}$. 
Networks from the ER ensemble 
have the smallest diameter followed by networks from the ER ensemble 
with fixed number of links in this small $P_{\rm B}$ region. 
Networks from the spatial network ensemble 
reveal the largest diameters also for large backup capacities.
Interestingly, graphs from the ER ensemble have a quite large 
diameter for larger backup capacities, whereas for smaller $P_{\rm B}$ the 
diameters are the smallest of all ensembles.
These results are somehow opposite to what was found in \cite{Hartmann14}, 
where networks from the spatial network ensemble 
reveal the smallest diameter for small backup capacities.

Networks from the ER ensemble with fixed number of edges show 
a decrease of the average diameter above $P_{\rm B} \approx 45$ 
and a small increase for very large 
backup capacities. This shows that the diameter is only a good observable 
to determine the resilience of a power flow network, if the edge number 
is flexible. Clearly, for real-world situations, where
the number of links is an economic factor, one aims at
minimal or at least constant edge number. Thus, 
for network ensembles with fixed number of edges other 
quantities may be considered.

One quantity is, e.g., what we call \textit{power sign ratio} 
(cf., \cite{Brede08,Kelly11}, where a \textit{frequency sign ratio} is 
used to characterize synchrony optimized networks). This quantity measures 
the fraction of links that connect 
synchronous machines whose power $P_i$ has opposite sign compared to the 
total number of edges in the network. In \fref{power_sign_ratio} the 
average of this quantity is shown for the studied network ensembles. The 
averaging is performed in the same way as for the graph diameter. For 
small $P_{\rm B}$ the power sign ratio is close to 0.5 for all network 
types. This means that on average from any two edges in the graph  
one of them connects machines with opposite signs of the power. 
This corresponds to the purely random case, since half
of the nodes exhibit positive and half of the nodes negative power.
Clearly, for increasing backup capacity, i.e., 
decreasing resilience $p_{-}$ also decreases. For 
networks from the spatial network ensemble this 
decrease is the most shallow.

Similarly to \cite{Brede08,Kelly11}, where a increasing value of the 
frequency sign ratio $p_{-}$ indicates enhancement of synchrony, here, 
$p_{-}$ (with the powers $P_i$) serves as a good indicator for the 
resilience of power flow networks.

\section{Summary and outlook}
We studied the resilience of power-flow models on networks against the 
failure of a transmission line. Three different random network ensembles, 
namely ER, spatial, and ER networks with fixed number of edges and 
the topology of the UK power grid were analyzed. The key quantity to 
determine the resilience of a network is the \textit{backup capacity}. 
It is defined by the additional capacity of the links which needs to be 
provided for stable operation in case of a failure of the link with the 
highest power flow. 
This quantity is a realistic measure of resilience, since
a power-grid blackout is very costly and should be avoided with a
large effort.
With a specific reweighting procedure the tails of 
the pdf below densities of $10^{-160}$ could be 
investigated. This allows for the study of very resilient and very 
vulnerable networks as well as typical ones.
In addition, the $p$-value allows for the comparison of a given 
network with an network ensemble by giving a quality measure for the 
investigated network.

For the UK power grid we found a typical backup capacity 
$P_{\rm B} \approx 1.3$ which is located in the right peak of the 
corresponding pdf generated for an ER ensemble with fixed 
number of links.
A $p$-value of 0.67 of the UK grid is indicating that it is of low 
significance with regard to resilience and that there exist many networks 
in the ER ensemble with fixed number of links that are more resilient.
The position of the two peaks in the pdf increase logarithmically with 
growing $N$ for the three ensembles of random networks.
The right tail of the pdfs for these three ensembles towards 
larger backup capacities is an exponential in 
about the left half of its support followed by a strong 
curvature in the right half of the support. 
This is confirmed by the rate function which converges 
to a corresponding limiting curve for increasing $N$.

Adding more links to a network makes it 
\emph{typically} more resilient, which is not surprising. 
Also, in the non-spatial ER ensembles, which allow for more freedom 
when placing the edges, it is easier to find resilient networks. 
Nevertheless, for real
applications, the two-dimensional model is more appropriate,
in particular since it is almost as likely as for the ER ensembles 
to find very resilient networks.
For this case more
interestingly, resilient networks are characterized by a small diameter 
and a large power sign ratio even for the ER ensemble with fixed number 
of links. The latter observation is quite interesting,
it means that power producers should be placed close to power consumers.
This is convenient since this strategy reduces the costs
for creating the network for transporting the electric power, as
it is classically done anyway. 
Thus, minimizing the transportation costs
and making the networks resilient are to a large extent 
not conflicting goals.

When using the $p$-value calculation, one should choose a 
suitable network ensemble for comparison. The ensemble should match the 
constraints of the investigated real-world network. 
Here, we used an ER ensemble with fixed number 
of edges for comparison with the UK grid as illustrating example.
For practical evaluations of existing or planned power grids one would 
include, e.g., 
geographical constraints or cost minimization. Within such a constrained 
ensemble the backup capacity of an existing grid would be located in 
the low-probability tail of the pdf. Thus, a large-deviation approach, 
like presented here, is necessary to evaluate such a power grid.

In the future, it would be interesting to investigate 
more thoroughly where the 
double-peaked structure of the pdf 
comes from. 
It might also be useful to consider more realistic, i.e., dynamic,
networks for electric power grids, as mentioned already above. In 
addition, one could use a spatial network ensemble which takes the 
costs of adding a transmission line into account to get an economically 
more realistic model.

\ack
The authors acknowledge valuable discussions with M.\ Timme, 
D.\ Witthaut, and B.\ Werther. 
Financial support was obtained by the Lower Saxony research network 
``Smart Nord'' which acknowledges the support of the Lower Saxony Ministry 
of Science and Culture through the ``Niedersächsisches Vorab'' grant 
programme (grant ZN2764/ZN 2896). The simulations were performed at the 
HERO cluster of the Carl von Ossietzky Universit\"at Oldenburg funded by 
the DFG through its Major Research Instrumentation Programme 
(INST 184/108-1 FUGG) and the Ministry of Science and Culture (MWK) of the 
Lower Saxony State. 

\section*{References}
\bibliography{LD_PG.bib}

\end{document}